\documentclass[12pt, twocolumn,superscriptaddress,english]{iopart}
\usepackage[utf8]{inputenc}

\usepackage{amssymb}
\usepackage{graphicx}
\usepackage{xcolor}
\usepackage{siunitx}
\usepackage{lipsum}

\usepackage{lineno}

\usepackage{hyperref}

\newcommand{\Tc}{$T_\text{c}$}
\newcommand{\ep}{$ep$}

\newcommand{\sh}{H$_3$S}

\makeatletter
\def\@fnsymbol#1{\ensuremath{\ifcase#1\or \dagger\or *\or \ddagger\or
   \mathsection\or \mathparagraph\or \|\or **\or \dagger\dagger
   \or \ddagger\ddagger \else\@ctrerr\fi}}
\makeatother

\begin{document}

\title{Phase diagram and superconductivity of Calcium Alanates under pressure}

\author{Simone Di Cataldo$^{1,2}$}
\author{Lilia Boeri$^{2,3}$}

\address{$^{1}$ Institut f\"{u}r Festk\"{o}rperphysik, Wiedner Hauptstraße 8-10, 1040 Wien, Austria}
\address{$^{2}$ Dipartimento di Fisica, Sapienza Universit\`a di Roma, 00185 Roma, Italy} 
\address{$^{3}$ Centro Ricerche Enrico Fermi, Via Panisperna 89 A, 00184 Rome, Italy}
\eads{\mailto{simone.cataldo@tuwien.ac.at}, \mailto{lilia.boeri@uniroma1.it}}

\date{\today}
\begin{abstract}
In this paper we present a first-principles study of the high-pressure superconducting phase diagram of calcium alanates (Ca-Al-H), 
based on ab-initio crystal structure prediction and anisotropic Migdal-Eliashberg Theory.
Calcium alanates have been intensively studied at ambient pressure for their
hydrogen-storage properties, but their high-pressure behavior is largely unknown.
By performing a full scan of the ternary convex hull at several pressures 
between 0 and 300 GPa, we identify several new structural motifs, characterized by
a high Al-H coordination, where Al $d$ orbitals participate in the bonding.
Among all new phases thus identified, we focus in particular on a phase with CaAlH$_7$ composition, which lies on the convex hull at  300 GPa, and remains dynamically stable down to 50 GPa, with a predicted superconducting \Tc{} of 88 K, which likely represents a new promising template to achieve increase chemical precompression in ternary hydrides.
Our findings reveal important insights into the structure-property relationships of calcium alanates under high pressure, and highlight a possible strategy to achieve conventional superconductivity at low pressures.
\end{abstract}


\submitto{\JPCM}

\maketitle
\ioptwocol
\begin{figure*}[htb]
\centering
	\includegraphics[width=1.05\columnwidth]{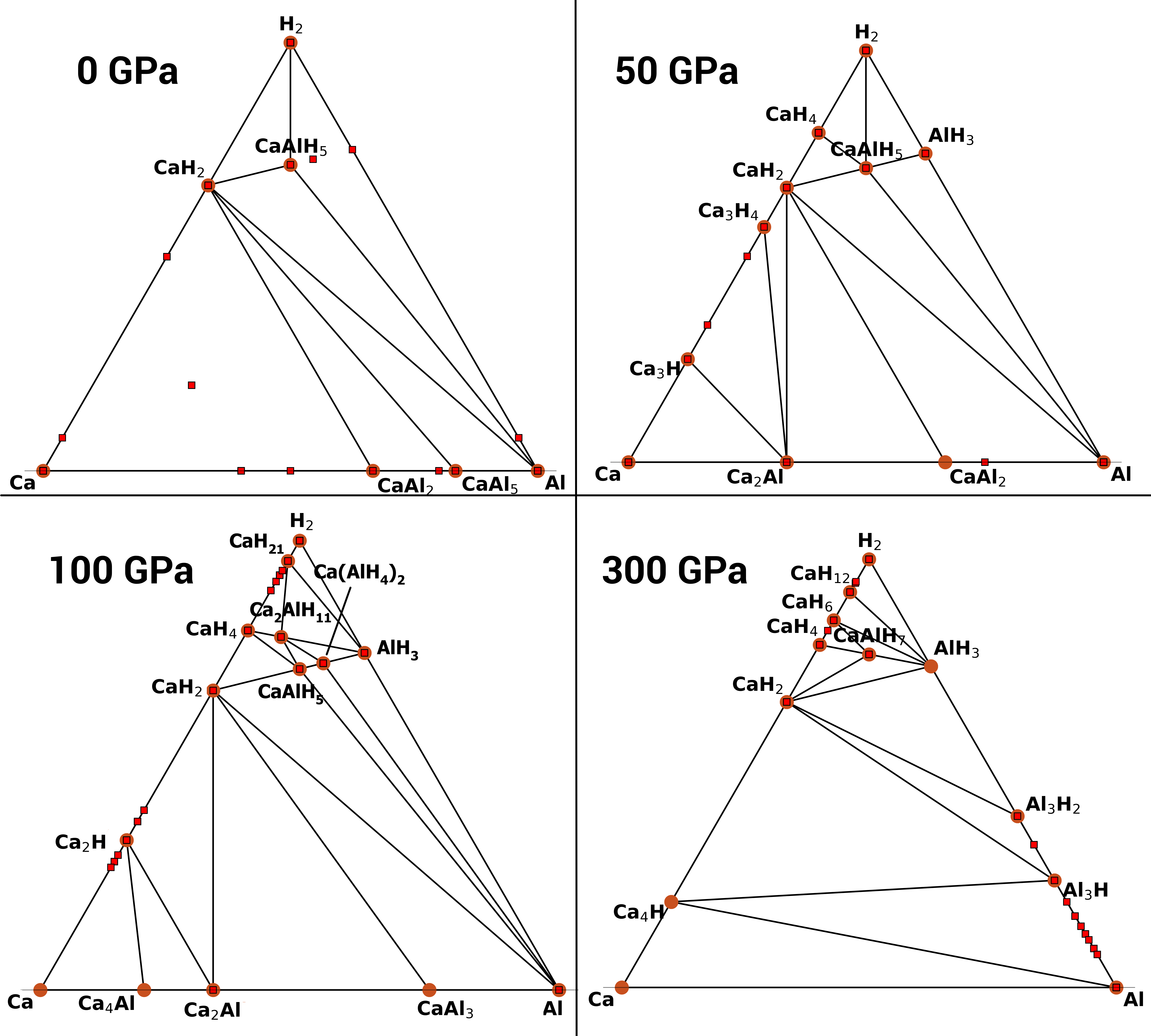}
	\includegraphics[width=0.85\columnwidth]{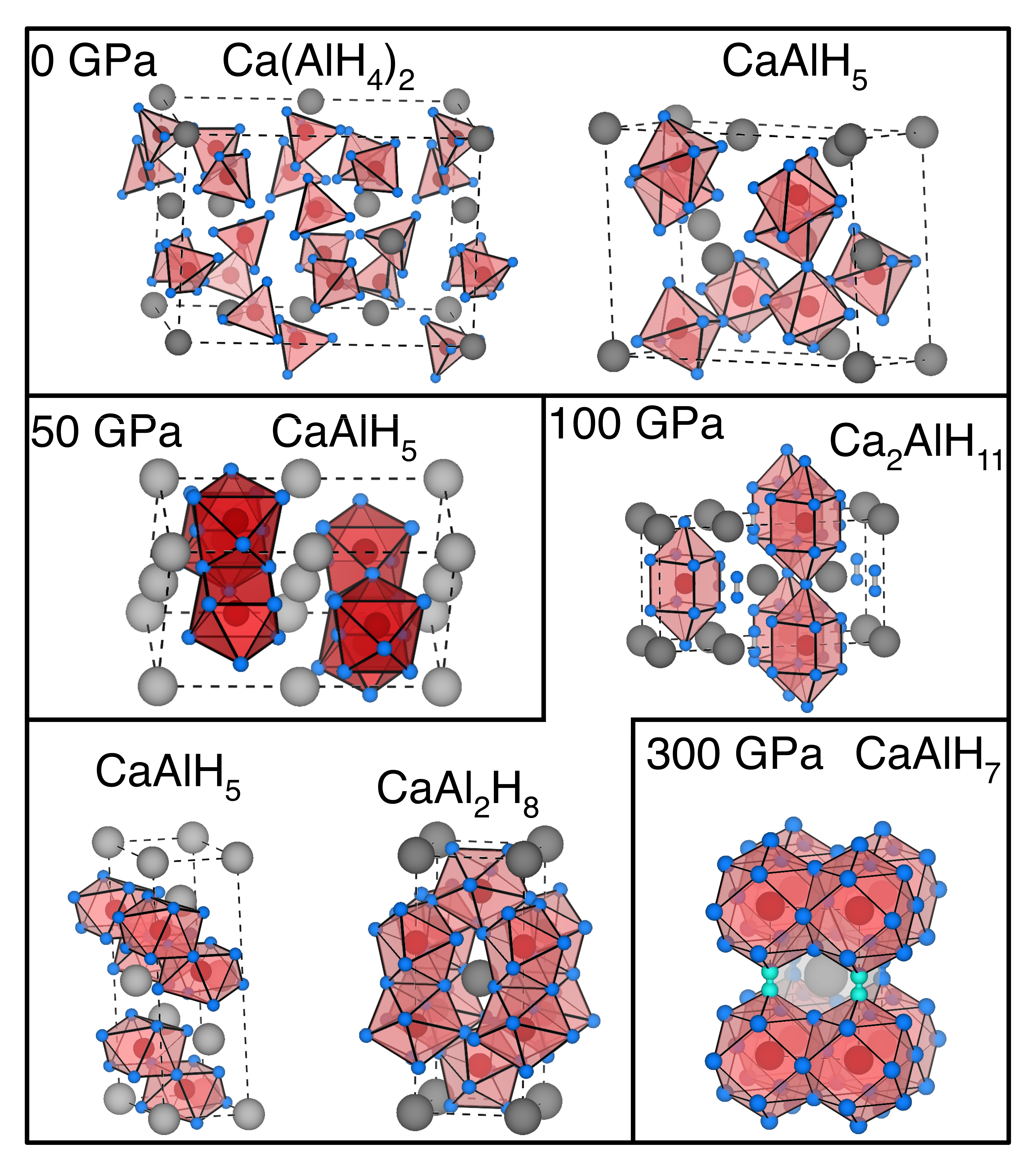}
	\caption{Convex hull diagrams for the Ca-Al-H system at 0, 50, 100 and 300 GPa. Thermodynamically stable compositions are indicated as orange circles. Compositions within 25 meV/atom of the hull are shown as red squares. Crystal structures for all thermodynamically stable ternary alanates at 0, 50, 100, and 300 GPa. Ca, Al, and H atoms are shown as grey, red, and blue spheres, respectively.}
 	\label{fig:figure1}
\end{figure*}

\section{Introduction}
The discovery of high-temperature superconductivity at 203 K in \sh{} in 2014 \cite{Eremets_Nature_2015_SH3, Eremets_NatPhys_2016_SH3} at Megabar pressures brought hydrides into the spotlight of superconducting materials research. Their extremely high \Tc{}s deriving from an electron-phonon mechanism finally demonstrated that conventional superconductors can, in fact, achieve high \Tc{}, contrary to previous misconceptions.

In this context, computational predictions based on ab-initio calculations have become an invaluable tool for materials discovery, often anticipating and guiding experiments towards the most promising materials \cite{Pickard_AnnRevCMP_2020_review, Boeri_PhysRep_2020_review, Boeri_JPCM_2021_roadmap}.
In the span of just eight years, all possible combinations of a single element plus hydrogen (binary hydrides) have been computationally explored \cite{Boeri_PhysRep_2020_review, Pickard_AnnRevCMP_2020_review, Oganov_NatRevMat_2019_str_pred_discovery} seeking novel high-temperature superconductors, several of which were also experimentally confirmed \cite{Hemley_PRB_2018_LaH, Hemley_PRL_2019_LaH, Eremets_Nature_2019_LaH, Oganov_MatToday_2020_ThH, Oganov_AdvMat_2021_YH6, Eremets_arXiv_2018_LaH10, Oganov_PRL_2021_CeH}. Computational studies of high-pressure hydrides  not only permitted to
identify unknown materials, but also to gain a much deeper understanding of the relationship between chemical bonding and conventional superconductivity \cite{Pickett_PRL_2001_MgB2, Mazin_PRB_2015_SH3, Heil_PRB_2015_bondingSH3, Heil_PRB_2019_YHx, Ma_PNAS_2012_CaHx_hull}, instrumental to the design of new materials with improved superconducting properties. 

In the last three years, the focus of hydride research has shifted towards identifying materials which can form and remain stable at more accessible pressure than record superhydrides, even at the cost of a reduced \Tc{}, as 
 this would open up the possibility of more practical applications. One possible route is to explore \textit{ternary} hydrides, i.e. compounds that contain 
 hydrogen and two other elements.
 Indeed, the presence of two different elements permits to
 realize a much wider variety of chemical environments for hydrogen. For example, we have shown that in lanthanum hydrides the addition of a third element stabilizes a high-T$_c$ structure with LaBH$_8$ composition down to 35 GPa \cite{DiCataldo_PRB_2021_LaBH8, Liang_PRB_2021_LaBH, Zhang_PRL_2022_LaXH}. We later demonstrated that the stabilization pressure could be further reduced with a careful choice of the elements in the La/B site within the same structural template
 down to 3 GPa in BaSiH$_8$. \cite{Lucrezi_NPJ_2022_BaSiH,  Lucrezi_arXiv_2023_QuantDyn}
 It is very likely that other mechanism of increased chemical precompression may be identified in other ternary hydrides,
 which offer an unexplored playground of more than 7000 potential combinations.

In this study, we focus on calcium alanates (CAH), 
a class of widely-available materials which has been extensively investigated at 
ambient pressure because of their hydrogen storage properties \cite{Peles_PRB_2004_alanates_d, Wolverton_PRB_2007_CaAlH, Huan_PRL_2013_alanates, Milanese_Metals_2018_review_alanates}. CAH are also closely related to calcium borohydrides (CBH), which we studied in a previous publication \cite{DiCataldo_PRB_2020_CaBH}, without finding viable candidates for increased chemical precompression.

At ambient pressure, both CBH and CAH form hydrogen-rich molecular crystals containing Ca$^{++}$ and $(YH_{x})_{2}^{-}$ anions ($Y$ = B, Al; $x$ = 4 for Al, $x$ = 2, 3, 4 for B) \cite{Ronnebro_JPCB_2007_CaBH_review, Milanese_Metals_2018_review_alanates}, which can absorb/desorb large amounts of hydrogen.
Despite sharing the same valence, Al differs from B because of the presence of empty $3d$ orbitals, which lie close in energy to the $2p$ states, and can participate in the bonding. 
Already at ambient pressure, partial occupation Al-$d$ orbitals stabilize a CaAlH$_5$ phase with corner-sharing octahedra \cite{Peles_PRB_2004_alanates_d, Wolverton_PRB_2007_CaAlH}, which is absent in the phase diagram of CBH. Since it is well known that high pressures 
(\textit{forbidden chemistry}) phases, i.e. phases with unusual compositions and configurations, particularly 
for elements with low-lying unoccupied orbitals,
we expect that also in CAH Al-$d$ orbitals will play an increasing role in the bonding, leading to a phase diagram
substantially different from that of CBH. This is indeed confirmed by our calculations, which show that CAH tends to form very complex high-pressure phases, with high Al-H coordination.

In particular, we identify a high-\Tc{} (82 K) CaAlH$_{7}$ phase, which should form at high pressures and remain dynamically stable down to 50 GPa, with an unusual structural template, in which planes of H-cages are connected by H-H bonds.
It is very likely that also this new $XY$H$_7$
could be further optimized by a careful substitution of other elements in place of calcium, leading to higher \Tc{} and/or lower stabilization pressures.

The paper is structured as follows: first we describe the ternary phase diagram at 0, 50, 100, and 300 GPa, and describe the thermodynamically stable structures. Second, we discuss the electronic properties of those structures. Third, we compare the low- and high-pressure structures of CAH and CBH. Finally, we discuss in detail the superconducting properties of the high-pressure CaAlH$_{7}$ structure.

\begin{figure*}[htb]
	\includegraphics[width=1.95\columnwidth]{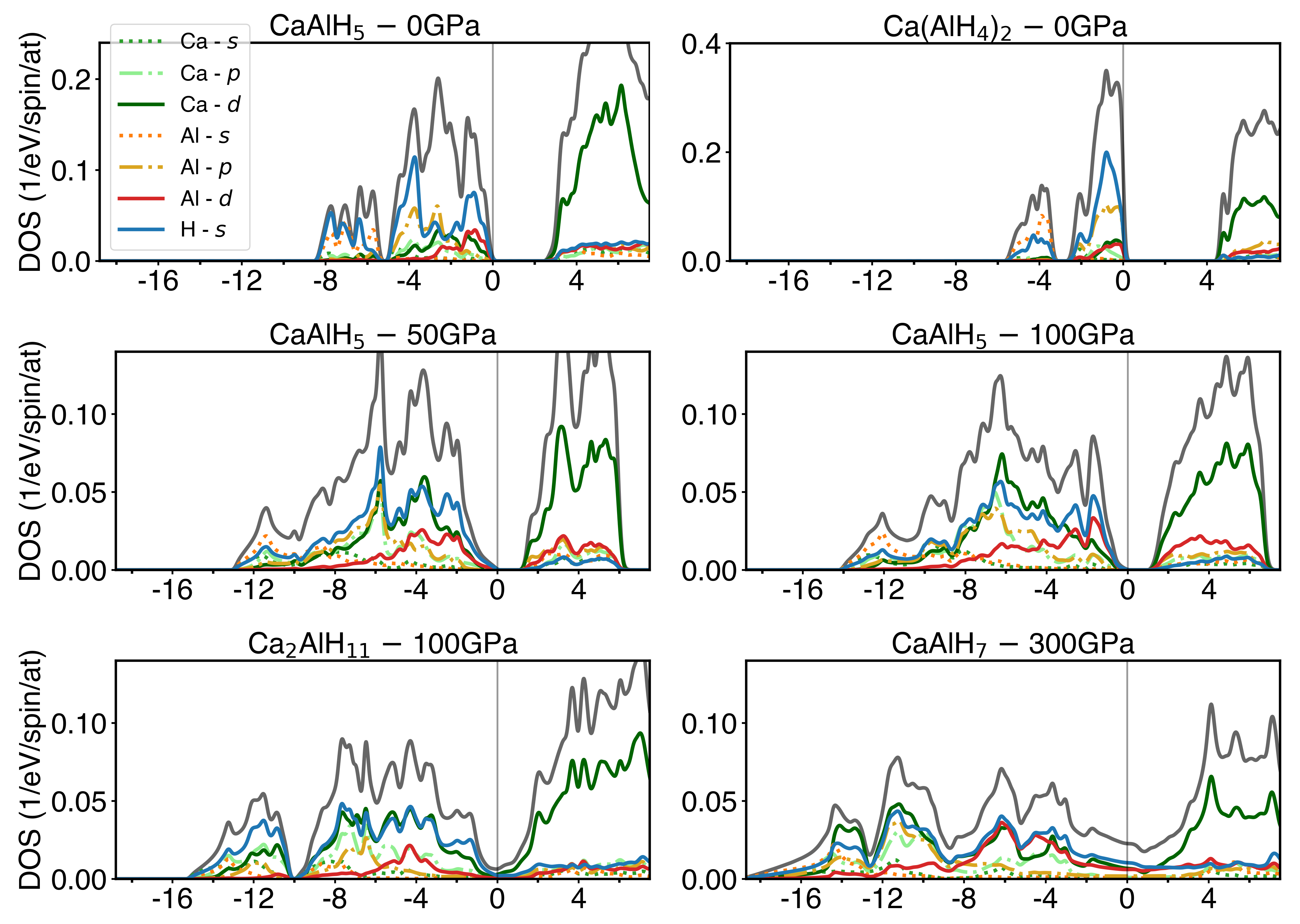}
	\caption{Electronic band structure for four selected calcium alanates. Black and colored lines indicate the DFT and Wannierized bands, respectively. The Wannier orbital onto which the band projection is carried out is indicated on the right side of each figure.}
	\label{fig:figure2}
\end{figure*}

\section{Results and discussion}
\label{sect:resultsdiscussion}

\subsection{Phase Diagram}
The phase diagram of CAH as a function of pressure was 
obtained computing the ternary convex hull,  
using \textit{ab-initio} evolutionary crystal structure prediction as implemented in the \textit{USPEX} code ~\cite{USPEX_1, USPEX_2}. For the underlying total energy calculations and relaxations
we employed the DFT Vienna ab-initio simulation package (VASP) \cite{Kresse_PRB_1996_VASP}, with Projector-Augmented Wave pseudopotentials \cite{Kresse_PRB_1999_VASP_pseudo} and PBE exchange-correlation functional.
Further details can be found in the SM. 

The left panel of figure \ref{fig:figure1} shows the convex hulls obtained at 0, 50, 100, and 300 GPa, for each of these pressures, we sampled over 5000 structures and over a hundred unique compositions. 
These values of pressures were chosen to ensure a reasonable sampling of low and intermediate/high-pressure phases, based on our previous experience on binary and ternary hydrides.
The right panel of the figure shows the crystal structures of the phases corresponding to stable compositions.

At ambient pressure (0 GPa) we find that the most stable ternary composition is CaAlH$_5$, although Ca(AlH$_4$)$_2$ is only 22 meV/atom above the hull, in agreement with calculations from Ref. \cite{Wolverton_PRB_2007_CaAlH}, as well as experiment, since both phases can be synthesized \cite{Iosub_HydrEnerg_2009_CaAlH5_synthesis, Mamatha_JAlloysComps_2006_CaAlH42_synthesys}.
The two structures are characterized by a qualitatively different geometry of the Al-H bonds: corner-sharing AlH$_{6}$ octahedra in CaAlH$_5$, and AlH$_{4}$ tetrahedra in Ca(AlH$_4$)$_2$, in both cases with interstitial calcium atoms. 

At 50 GPa, CaAlH$_5$ remains the only stable composition, and the stable structure contains face- and corner-sharing AlH$_{8}$ polyhedra with square antiprismatic geometries, alternating with Ca in a body-centered orthorhombic sublattice. 

At 100 GPa, the stable compositions are CaAlH$_5$, CaAl$_2$H$_8$, and Ca$_2$AlH$_{11}$. The strucuture of CaAlH$_5$ contains the same AlH$_8$ square antiprisms seen at 50 GPa, but now arranged in a corner- and edge-sharing pattern which compenetrates the calcium sublattice. In Ca(AlH$_4$)$_2$,
 which re-enters as a stable composition, the structure is radically different from the ambient pressure one, as it shows a sublattice of corner- and edge-sharing AlH$_8$ distorted snub disphenoids encaging Ca atoms. The ground-state structure of Ca$_2$AlH$_{11}$ is characterized by a lattice of AlH$_{10}$ elongated square bipyramids which sharing the top vertex, alternated with interstitial Ca atoms and trapped H$_2$ molecules.

Finally, at 300 GPa the only stable ternary composition is CaAlH$_{7}$. The crystal structure contains a combination of side-sharing AlH$_{12}$ cuboctahedra which share faces with CaH$_{16}$ truncated cubes capped with square pyramids. The two combined polyhedra fully tessellate the space. The AlH$_{12}$ cuboctahedra lie in separate planes, connected by a H--H bond, highlighted by green hydrogen atoms in Fig. \ref{fig:figure1} (right panel). The H-H bond distance increases with increasing pressure, going from 0.86 to 0.95 $\AA$ from 50 to 300 GPa. This value indicates delocalized atomic H-H bonds rather than molecular ones.

Overall, the structural changes in ternary calcium alanates (CAH) with increasing pressure suggest a profound change in the orbital hybridization. At ambient pressure, the presence of octahedral AlH$_{6}$ motifs indicates that H partially hybridizes with Al--$d$ states, whereas the AlH$_{4} ^{-}$ motifs indicates H--Al $sp^{3}$ hybridization. The tetrahedral geometry is also observed in Ca(BH$_4$)$_2$, while stable Ca(BH$_3$)$_2$ and Ca(BH$_2$)$_2$ compositions correspond to $sp^{2}$ and $sp$ hybridization \cite{Wolverton_PRB_2010_CaBHx, DiCataldo_PRB_2020_CaBH}. 

As pressure increases, however, CAH behaves in a substantially different way from CBH, as Al--$d$ states are effectively
pulled down in energy compared to $s,p$ states.
In the CBH phase diagram only structures with $sp$, $sp^2$, and $sp^3$ hybridization are stable or weakly metastable up to 150 GPa, and some  survive up to 300 GPa; even at 300 GPa, the B-H coordination is never larger than six. 
In CAH, already at 50 GPa, stable structures exhibit complex polyhedral structures; the average number of vertices and faces increases with pressure, up to a 12-coordinated Al-H polyhedron in CaAlH$_{7}$.

\subsection{Electronic properties}

The trend in crystal structures suggests that in many CAH structures Al-$d$ orbitals participate in the chemical bond with H. To make the argument less qualitative, in Fig. \ref{fig:figure2} we show the partial density of states (pDOS) \cite{pdos_note}, calculated for all the thermodynamically stable structures: Ca(AlH$_4$)$_2$ and CaAlH$_5$ at 0 GPa, CaAlH$_5$ at 50 GPa, CaAlH$_5$ and Ca$_2$AlH$_{11}$ at 100 GPa, and CaAlH$_{7}$ at 300 GPa. Note that in the Figure the Fermi energy (for metals) and the valence band maximum (for insulators) is taken as zero. 

In CaAlH$_5$ at 0 GPa, the DOS in the valence region is characterized by two peaks; the electronic states can be understood in terms of molecular AlH$_{6}$ octahedra, resulting from Al-$sp^{3}d^{2}$ hybridization. The system is an insulator with a wide bonding/antibonding gap of 2.6 eV. 
The DOS in the valence region of Ca(AlH$_4$)$_2$ is characterized by two well-separated peaks, extending from about -6 to -3 and to -3 to 0 eV. These states exhibit a very small dispersion, and like in Ca(BH$_4$)$_2$ \cite{DiCataldo_PRB_2020_CaBH}, correspond to $sp^{3}$ molecular orbitals of the AlH$_{4}^{-}$ anion. We note, however, that a small projection of Al-$d$ states is found near the top of the valence band. This system is also insulating, with a wide boding/antibonding gap of 4.5 eV, in line with other calculated values in the literature \cite{Jensen_SSHydrStorage_2008_review, Jensen_ChemSocRev_2017_MBH}.

 CaAlH$_5$ at 50 GPa is still insulating, but the gap is reduced to 1.2 eV. Here the occupied states merge into a single peak, and the weight of Al-$d$ states is significantly enhanced compared to lower pressures. 

At 100 GPa, the electronic structure of CaAlH$_5$ is very similar to the one at 50 GPa, with a gap only slightly reduced to about 1 eV.  Ca$_2$AlH$_{11}$, on the other hand, is a compensated semimetal. 

The behavior of the only phase stable at 300 GPa is  qualitatively very different from what observed at lower pressures. CaAlH$_7$ is metallic, with strongly dispersed electronic bands. Here, Al-$d$ states give a non-zero contribution to the DOS down to the bottom of the valence band at -15 eV, indicating a major rearrangment of the bonds.

\begin{figure}[h]
	\includegraphics[width=0.99\columnwidth]{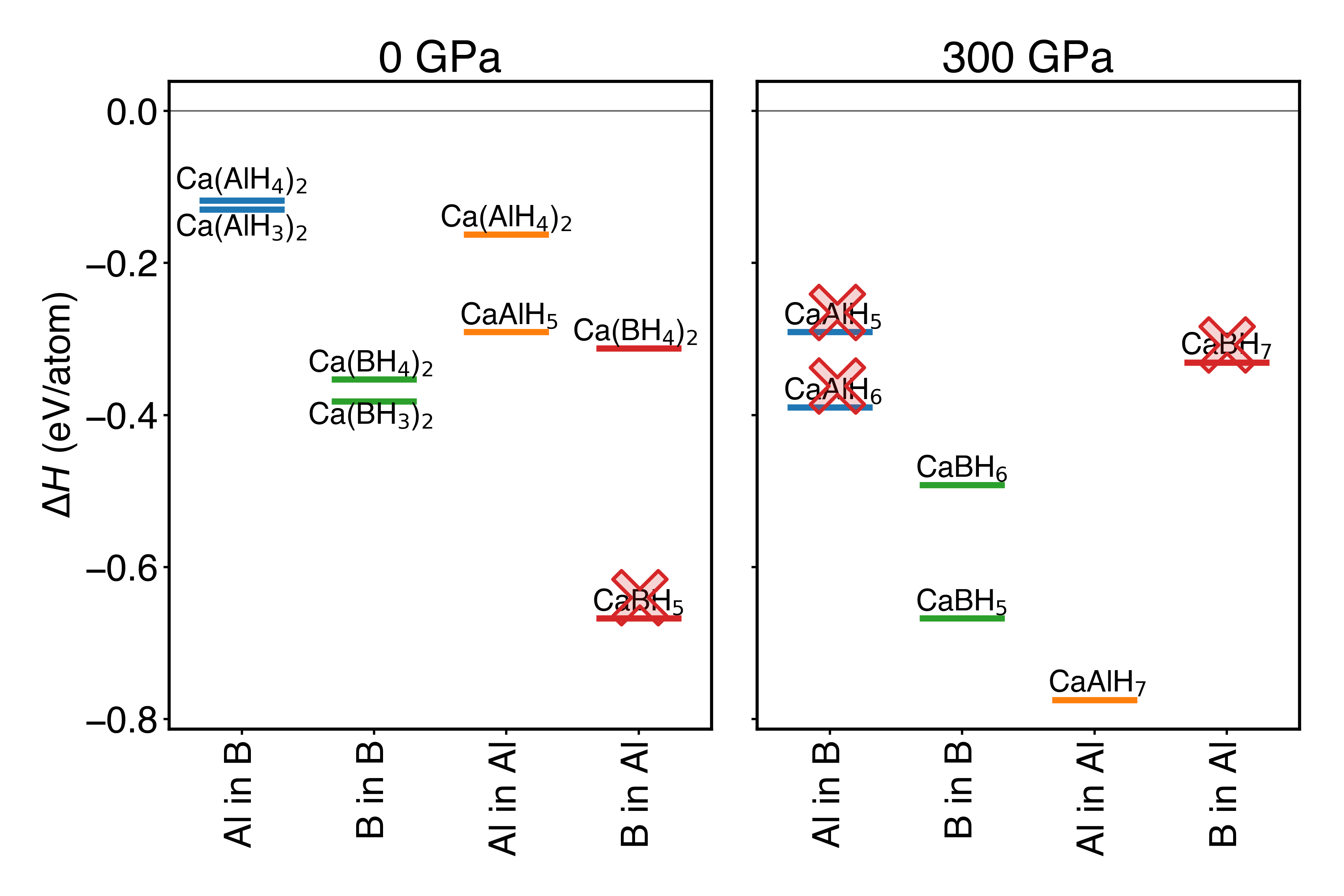}
	\caption{Formation enthalpy ranking of different structures with atomic substitution of B (Al) into the Al (B) site. The formation enthalpy is calculated with respect to pure elements. The formation enthalpy is indicated by a colored line, along with the composition. The substitution is indicated by the labels over the $x$ axis (e.g. an Al substitution in the B site is indicated by \textit{Al in B}). Structures that are dynamically unstable or relax into a different phase are indicated with a red cross.}
	\label{fig:figure3}
\end{figure}

In order to further elucidate the increasing role of 
Al--$d$ orbitals in the bonding, we performed an additional analysis, in which we 
we performed a systematic substitution of Al in CBH structures and vice versa, and evaluated the thermodynamical and dynamical stability of the resulting structures. 
In Fig. \ref{fig:figure3} we show the ranking of the resulting structures according to formation enthalpy \footnote{The formation enthalpy is considered with respect to the pure elements, e.g. $\Delta H\left(CaAlH_{5}\right) = H\left(CaAlH_{5}\right) - H\left(Ca\right) - H\left(Al\right) - \frac{5}{2}H\left(H_{2}\right)$}. 

When Al is substituted in the ambient-pressure structures of B (Ca(BH$_3$)$_2$ and Ca(BH$_4$)$_2$), the relaxed structures retains the same qualitative features as the original structure, and turn out to be competitive in energy with the lowest-energy structure for Ca(AlH$_4$)$_2$.\footnote{The data on the crystal structures after the relaxation is available as a compressed file in the Supplementary Material \cite{suppmat}}.
This can be easily understood, as Al employs in this structure its $s,p$ valence orbitals.
The $B \to Al$ substitution is, on the other hand, more problematic: while Ca(AlH$_4$)$_2$) also retains its qualitative features upon B substitution, CaAlH$_5$ does not,
and the structure "breaks down" upon relaxation, giving rise to a distorted combination of BH$_4$ anions and atomic hydrogen. In fact, the B 3$d$ orbitals lie too far away in energy from the $2s,2p$ valence orbitals to participate in the bonding.

At 300 GPa, we substituted Al in the structures with CaBH$_5$ and CaBH$_6$ composition that we found to be stable for the Ca-B-H system at the same pressure in our previous work\cite{DiCataldo_PRB_2020_CaBH}. Both structures exhibit unusual B-H bonding: the former is characterized by BH$_5$ triangular bipyramids, the latter by BH$_6$ 6-vertex antiprisms. Al substitution in the CaBH$_5$ and CaBH$_6$ structures leads to a qualitative change in the crystal structure. In CaBH$_5$ the BH$_5$ triangular bipyramids drastically rearrange into side-sharing AlH$_{10}$ irregular polyhedra, while in CaBH$_6$ the 6-vertex antiprisms become corner-sharing regular cuboctahedra. The opposite process, i.e. B substitution in CaAlH$_7$ exhibits the same qualitative features, but the structure is dynamically unstable.


\begin{figure}[h]
	\includegraphics[width=0.95\columnwidth]{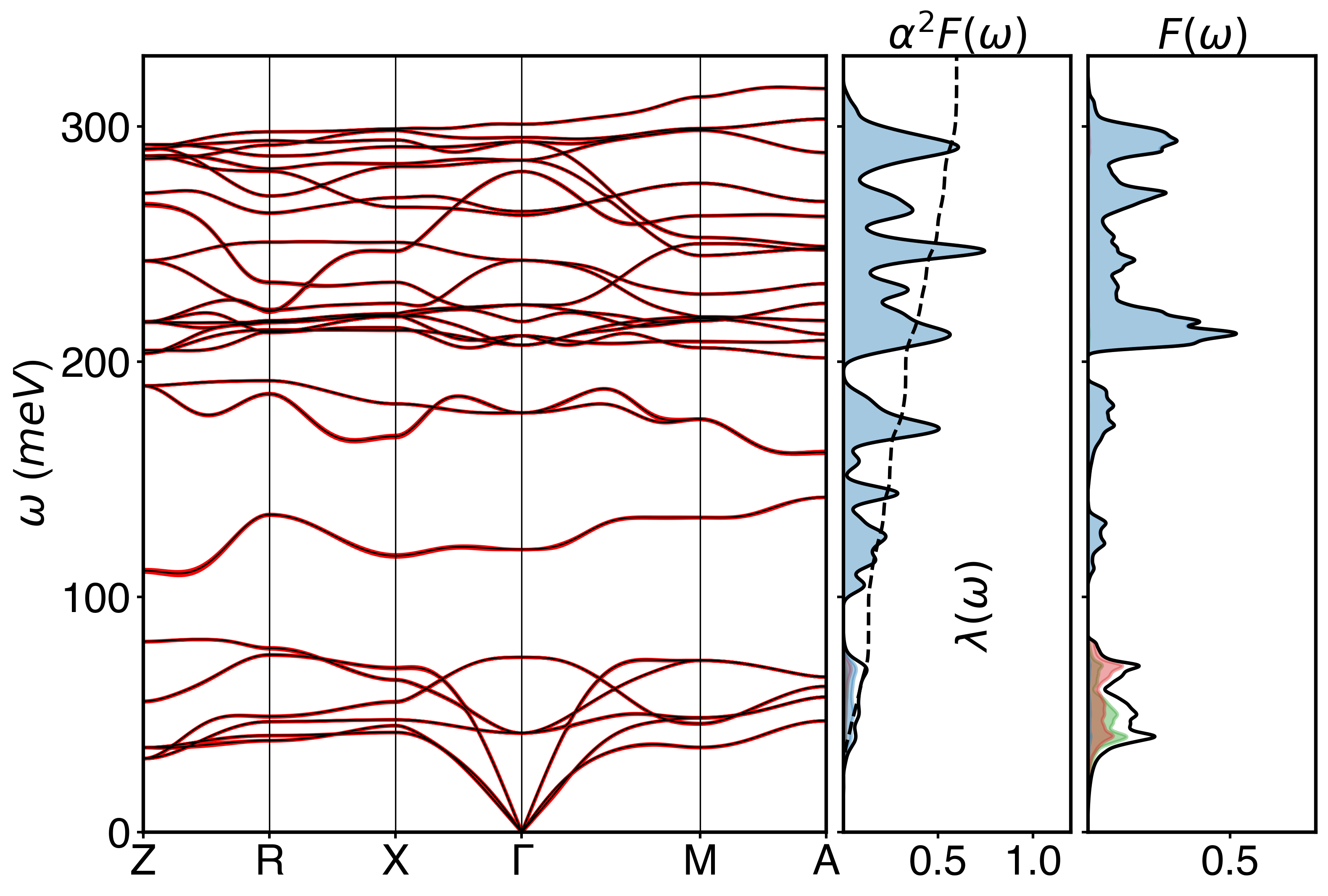}
	\caption{Phonon dispersions, atom-projected phonon density of states ($F(\omega)$) and \'{E}liashberg function ($\alpha^2F(\omega)$) for CaAlH$_{7}$ at 300 GPa. The total $F(\omega)$ and $\alpha^2F(\omega)$ are shown as solid black lines, while their projections onto Ca, Al, and H are shown as green, red, and orange filled curves, respectively.}
	\label{fig:figure4}
 	\includegraphics[width=0.95\columnwidth]{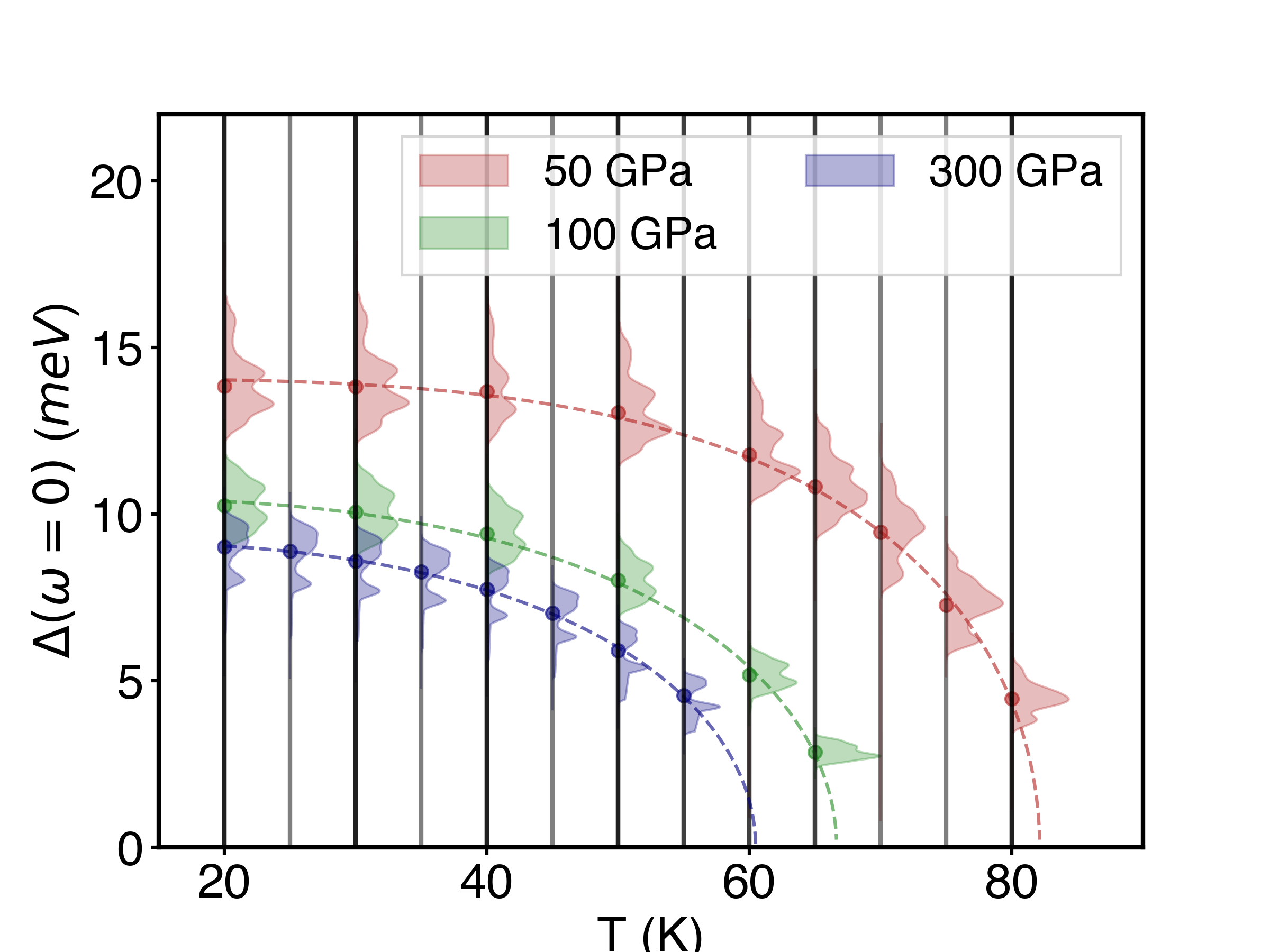}
	\caption{Leading edge of the anisotropic superconducting gap at 50 (red) and 300 GPa (blue) as a function of temperature. The corresponding superconducting \Tc{} is shown in Tab. \ref{tab:supercon}. The interpolating line is obtained from a fit with the function $\Delta(T) = \Delta_0 \sqrt{k \frac{T_c-T}{T}}$ of the weighted average of the anisotropic gap at $\omega = 0$.}
	\label{fig:figure5}
\end{figure}

\subsection{Superconductivity}
Of all the CAH structures predicted in this work, CaAlH$_{7}$ is the only one with the qualitative prerequisites to host high-\Tc{} superconductivity \cite{Errea_NatComm_2021_hydridebased, Boeri_PhysRep_2020_review}: a dense hydrogen sublattice, metallic behavior, and a significant fraction of hydrogen DOS at the Fermi level. We computed its electron-phonon coupling properties at 50, 100, and 300 GPa using Density Functional Perturbation Theory \cite{Savrasov_PRB_1996_ep_I, Baroni_RevModPhys_2001_DFPT, quantumespresso_1, quantumespresso_2}.

In Fig. \ref{fig:figure4} we show the phonon dispersions, Eliashberg function and phonon DOS at 300 GPa (The phonon dispersions at the other pressures are shown in the Supplementary Figure 3 \cite{suppmat}). Similarly to most hydrides, the total \ep{} coupling is spread over the whole optical branch, and the Eliashberg function has a predominantly hydrogen character, suggesting that the H-H 
intraband interactions, rather than Al-H interband ones, are providing most of the coupling. We observe a soft mode in the branch around 100 meV around the $X$ point, and another in the acoustic branch around the $R$ point (See also Supplementary Figure 3 \cite{suppmat}). The latter drives the system dynamical unstable below 50 GPa. Some hydrogen modes between 100 and 200 meV are not strongly dispersive, indicating that these modes correspond to short-range, molecular-like H-H vibrations, while between 200 and 300 meV the larger dispersion  indicates collective vibrations of the H cages, analogous to sodalite-like clathrate hydrides \cite{Ma_PNAS_2012_CaHx_hull, Heil_PRB_2019_YHx}.
We note that non-dispersive modes correspond to phonon eigenvectors parallel to the $c$ axis, and vice versa. 

Using the calculated \ep{} spectrum we calculated the superconducting \Tc{} 
by solving self-consistently  the anisotropic Migdal-Eliashberg equations on a fine interpolated grid \footnote{32 $\times$ 32 $\times$ 32 for both electrons and phonons} at 50, 100 and 300 GPa, using the \verb|EPW| code \cite{Giustino_PRB_2007_epw, Giustino_CPC_2016_EPW}, and a constant value of the Morel-Anderson pseudopotential $\mu^* = 0.10$, further details are available in the Supplementary Materials. 

In Fig. \ref{fig:figure5} we show the calculated leading edge of the superconducting gap at $\omega = 0$ for different pressures. The superconducting gap is quite isotropic, and  spreads  over a small energy interval of 2-3 meV.
(Further details on the anisotropic gap are provided in the Supplementary Material).

\begin{table*}[h]
    \centering
    \begin{tabular}{c|c|c|c|c|c}
     Comp.       &    P (GPa) &   $\lambda$   &   $\omega_{log} (meV)$    & T${c}^{MM}$ (K)  & $\frac{2 \Delta}{T_c}$ \\
     \hline
     \hline
    CaAlH$_{7}$      &  50       &      1.06    &        68         &   60  &  3.97 \\
    CaAlH$_{7}$      &  100       &       0.85    &        103      &   63  &  3.65 \\
    CaAlH$_{7}$      &  300       &        0.66     &       146     &   50  &  3.52 \\
    \end{tabular}
    \caption{Superconducting properties of the high-pressure CaAlH$_{7}$ phase at various pressures. The critical temperature is obtained by self-consistently solving the anisotropic \'{E}liashberg equations until the gap is converged within $10^{-2}$ meV (T${c}^{aniso}$), using a value of 0.10 for the Morel-Anderson pseudopotential $\mu^{*}$.}
    \label{tab:supercon}
\end{table*}

The \Tc{} is determined by fitting the average of the leading edge of the superconducting gap with an interpolating function (See caption of Fig. \ref{fig:figure5}), and extrapolating to $\Delta = 0$. The results are summarized in Tab. \ref{tab:supercon}. The ratio $\frac{2 \Delta}{T_c}$ is exactly 3.52 at 300 GPa, and it increases with decreasing pressure up to 3.97 at 50 GPa, as a soft-mode boosts coupling at lower pressure and pushes the system away from the weak-coupling limit.

The \Tc{} increases from 61 K at 300 GPa, to 82 K at 50 GPa, as a consequence of mode softening. In fact, the e-ph coupling constant $\lambda$ increases from 0.66 to 1.06, while  the logarithmic-average frequency $\omega_{log}$ decreases by almost one third (from 150 to 68 meV) in the same interval. 

A \Tc{} of 82 K at 50 GPa places CaAlH$_7$ in the same class
as XYH$_{8}$ ternary clathrate hydrides, \cite{DiCataldo_PRB_2021_LaBH8, Lucrezi_NPJ_2022_BaSiH, Zhang_PRL_2022_LaXH} i.e. that of ternary hydride superconductors, where efficient chemical precompression stabilizes a dense hydrogen sublattice at lower pressures
than binary hydrides. It is very likely that
\Tc{} and stabilization pressure may further be optimized by careful chemical substitution as was shown for LaBH$_8$ and BaSiH$_8$\cite{DiCataldo_PRB_2021_LaBH8, Lucrezi_NPJ_2022_BaSiH}.

\section{Conclusions}
\label{sect:conclusions}
In conclusion, using \textit{ab initio} methods based on Density Functional Theory, we studied the phase diagram and the superconducting properties of calcium aluminum hydrides (CAH) under pressures of 0, 50, 100, and 300 GPa. We found several stable phases in which aluminum progressively increases its coordination with hydrogen as pressure increases. 

In particular,  we find a structure with CaAlH$_{7}$ composition which, to the best of our knowledge, is still unreported. The structural motif comprises layers of AlH$_{12}$ cage-like polyhedra (cubooctahedra), linked by atomic H-H bonds, and is thus structurally analogous to
other ternary $XY$H$_n$ hydrides, where H cage-like units 
can be retained down to relatively low pressures, due to
the chemical precompression exerted by the X/Y sublattice. 

CaAlH$_{7}$ is thermodynamically stable at 300 GPa, but remains dynamically stable down to 50 GPa, where we predict a superconducting \Tc{} of 82 K by self-consistently solving the fully anisotropic Migdal-Eliashberg equations. 

 Hence, CaAlH$_{7}$ is one of the very few hydrides retaining
 high- \Tc{} superconducting properties down below Megbar pressures. Like in LaBH$_8$, where we have demonstrated the stabilization pressure can be lowered significantly by Ba/Si substitution, the CaAlH$_7$ structure could also be further optimized by replacing Ca with other alkaline metals or earths, and Al with other non-metals such as Ga, In, or Sn.



\section{Acknowledgments}
L.B. and S.D.C. acknowledge funding from the Austrian Science Fund (FWF) P30269-N36 and support from Fondo Ateneo-Sapienza 2018-2021. S.D.C. acknowledges computational resources from CINECA, proj. IsC90-HTS-TECH and IsC99-ACME-C, and the Vienna Scientific Cluster, proj. 71754 "TEST". LB acknowledges support from Project PE0000021, “Network 4 Energy Sustainable Transition – NEST”, funded  by the European Union – NextGenerationEU, under the National Recovery and Resilience Plan (NRRP), Mission 4 Component 2 Investment 1.3 - Call for tender No. 1561 of 11.10.2022 of Ministero dell’Universit\'{a} e della Ricerca (MUR).

\section*{References}
\bibliographystyle{iopart-num}

\begin{thebibliography}{10}
\bibitem{pdos_note}
The partial dos was obtained by projecting the total DOS onto spherical
  harmonics centered on each atom, and integrated over a radius equal to the
  atomic radius.
\end{thebibliography}

\providecommand{\newblock}{}

\end{document}